\documentclass[10pt,conference,letter,twocolumns]{ieeetran}

\pdfoutput = 1

\usepackage{graphicx}
\usepackage{color}
\usepackage{longtable}
\usepackage{amsfonts}
\usepackage{amsmath}
\usepackage{epsfig}
\usepackage{epstopdf}

\usepackage[utf8]{inputenc}
\newcommand{\fig}{Figure }

\newcommand{\specialcell}[2][c]{ \begin{tabular}[#1]{@{}c@{}}#2\end{tabular}}

\renewcommand{\baselinestretch}{1.05}
\begin{document}

\title{Multiband CSMA/CA with RTS-CTS Strategy}
\author{Baher \textsc{Mawlawi}$^{1,2,3}$, Jean-Baptiste \textsc{Doré}$^{1}$, Nikolai \textsc{LEBEDEV}$^{2,3,4}$, Jean-Marie \textsc{Gorce}$^{2,3}$\\
$^1$ CEA-Leti Minatec, 17 rue des Martyrs, 38054 Grenoble Cedex 9, France\\
$^2$University of Lyon, INRIA \\
$^3$INSA-Lyon, CITI-INRIA, F-69621, Villeurbanne, France \\
$^4$CPE Lyon,\ BP 2077, F-69616, France\\
\{baher.mawlawi,~jean-baptiste.dore\}@cea.fr\\
\{lebedev@cpe.fr,~jean-marie.gorce@insa-lyon.fr\}\\
}

\maketitle

\begin{abstract}
We present in this paper a new medium access control (MAC) scheme devoted to orthogonal frequency division multiple access (OFDMA) systems which aims at reducing collision probabilities during the channel request period. 
The proposed MAC relies on the classical carrier sense multiple access/collision avoidance (CSMA/CA) protocol with RTS / CTS ("Request To Send" / "Clear To Send") mechanism. The proposed method focus on the collision probability of RTS messages exploiting a multi-channel configuration for these messages while using the whole band for data transmissions. The protocol may be interpreted as an asynchronous frequency multiplexing of RTS messages. This method achieves strong performance gains in terms of throughput and latency especially in crowded networks.
\end{abstract}

\hfill

\section{Introduction}
In recent years, the fast increasing demand for high-speed wireless internet access motivated researchers to make efforts for improving the efficiency of decentralized wireless networks. The development of numerous new services on wireless terminals indeed lead to a strong expansion of the number of users causing an important deterioration of these networks in terms of throughput and system performance \cite{bianchi1} \cite{bianchi2}.

The traditional single band CSMA/CA system has the advantage of requiring neither signaling for bandwidth request nor planned allocation. However, its effectiveness degrades rapidly with the increasing number of simultaneous source nodes. This limitation can be overcome by using multiple division access on different bands where several source nodes can transmit simultaneously. Sources are familiar with the availability status of each band at each time instant. This multiple access on different bands may operates with OFDMA (Orthogonal Frequency Division Multiple Access) whereby the spectral resource (bandwidth) is divided into several orthogonal sub-carriers. This set of sub-carriers is further split into subsets, each subset constituting one communication channel. Source nodes then compete for accessing and sharing these resources in both time and frequency.

Different works already proposed to generalize the CSMA/CA to the multiband case \cite{av2} \cite{rawpeach} \cite{multichannel} in order to increase the global data rate. In these protocols, users are multiplexed through different channel while keeping the classical CSMA/CA strategy in each channel.

Other works tried to eliminate collisions between control and data packets by separating physically the control and the data planes: one band is reserved for control packets and the rest for data transmissions \cite{multi1} \cite{multi2} \cite{multi3}. This scheme provides a higher throughput compared to the classical protocol adopted in 802.11 standard. However, it suffers from two issues when the network is crowded or lightly busy. In crowded situations, the classical CSMA/CA still runs on a common channel and suffers from collisions between control messages. In low traffic conditions, high rates users are penalized because they cannot transmit simultaneously on several channels, even if several ones are free. 

We propose in this paper to adapt the CSMA/CA with RTS/CTS mechanism to address both issues. We prove by simulations that the outcome of the proposed protocol in terms of saturation throughput is better than the single band case and it remains quasi-constant for dense networks. The system delay is improved as well.

The paper is outlined as follows. We describe and justify the proposed protocol  in Section II and the system model is derived. Section III presents different scenarios exploiting the proposed protocol. Simulation results are presented in Section IV and the protocol performance is analysed. Finally, section V is reserved for conclusion.

\section{{System Model}}
\label{protocol}
As described in the introduction, we consider a CSMA/CA protocol with RTS/CTS scheme \cite{bianchi3}. Actually, the throughput is closely related to the collision rate between users \cite{basicpaper}. Considering an ideal channel, collisions may occur only during RTSs transmissions. Sending RTS on orthogonal bands may help to reduce drastically the collision probability. In this paper, we consider orthogonal frequency multiplexing for these RTS messages.

We consider a spectrum divided into $N$ bands. We assume that RTS messages have the same time duration for all users present in the network and that all transmitters (TX) and receivers (RX) have the knowledge of the band size and central positions of each band. We further assume that these nodes are able to work simultaneously on these bands which is made possible by the use of software radio transceivers.  

\begin{figure}[tb]
\begin{center}
\includegraphics[width=1\columnwidth]{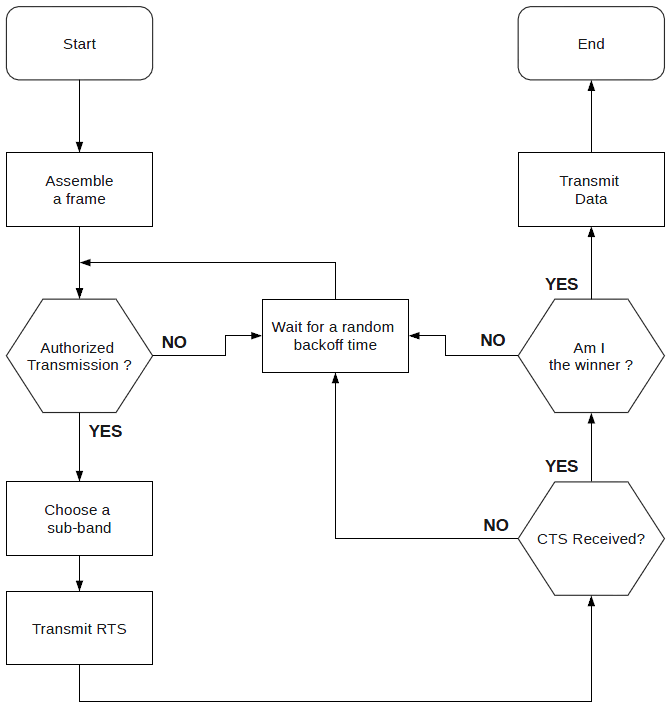}\\
\caption{Flow chart of the proposed protocol.}
\label{machine_etat}
\end{center}
\end{figure}

The proposed scheme is used to avoid collisions between multiple users (source nodes) requesting simultaneously an access to the channel. According to this protocol, a source node wishing to transmit data should first listen to the communication channel. A flow chart of the proposed protocol is depicted in Figure \ref{machine_etat}.

If the channel is busy, a period (expressed in number of time slots) of a waiting counter (known as "backoff counter") is chosen randomly in the interval [0, CW-1], where CW is a contention window. The channel is declared busy if there exists a signal on at least one band. The backoff counter is decremented by one each time the channel is detected to be available for a Distributed Inter-Frame Space (DIFS) duration. The wait counter freezes when the channel is busy, and resumes when the channel is available again for at least DIFS time.

When the backoff counter reaches zero, the source randomly chooses one band over the $N$ to send a permission request message (RTS) to the destination node. It waits for receiving an authorization message (CTS) from the destination node before transmitting data. The destination (AP) listens simultaneously all the bands. If one or more RTS is detected, the AP broadcasts a CTS message over the whole band, indicating which station is allowed to transmit. The bandwidth of CTS messages is $N$ times the bandwidth of RTS messages.

The chosen station (STA) sends its data and waits for Acknowledge (ACK) from the AP. Both data and ACK messages are sent using all the available bandwidth. Upon receipt of all transmitted data (successful transmission), and immediately, after a SIFS duration ("Short Inter-Frame Space"), the destination node sends an ACK (for "Acknowledgment"). The Contention window (CW) is an integer between CWmin and CWmax. CW is initially set to the minimum value: $CW=CW_{min}$. Whenever a source node is involved in a RTS collision, it increases the transmission waiting time by doubling the CW, up to the maximum value CWmax. Conversely, in the case of a successful RTS transmission, the source node reduces its CW to CWmin. 

\begin{figure}[tb]
\begin{center}
\includegraphics[width=1\columnwidth]{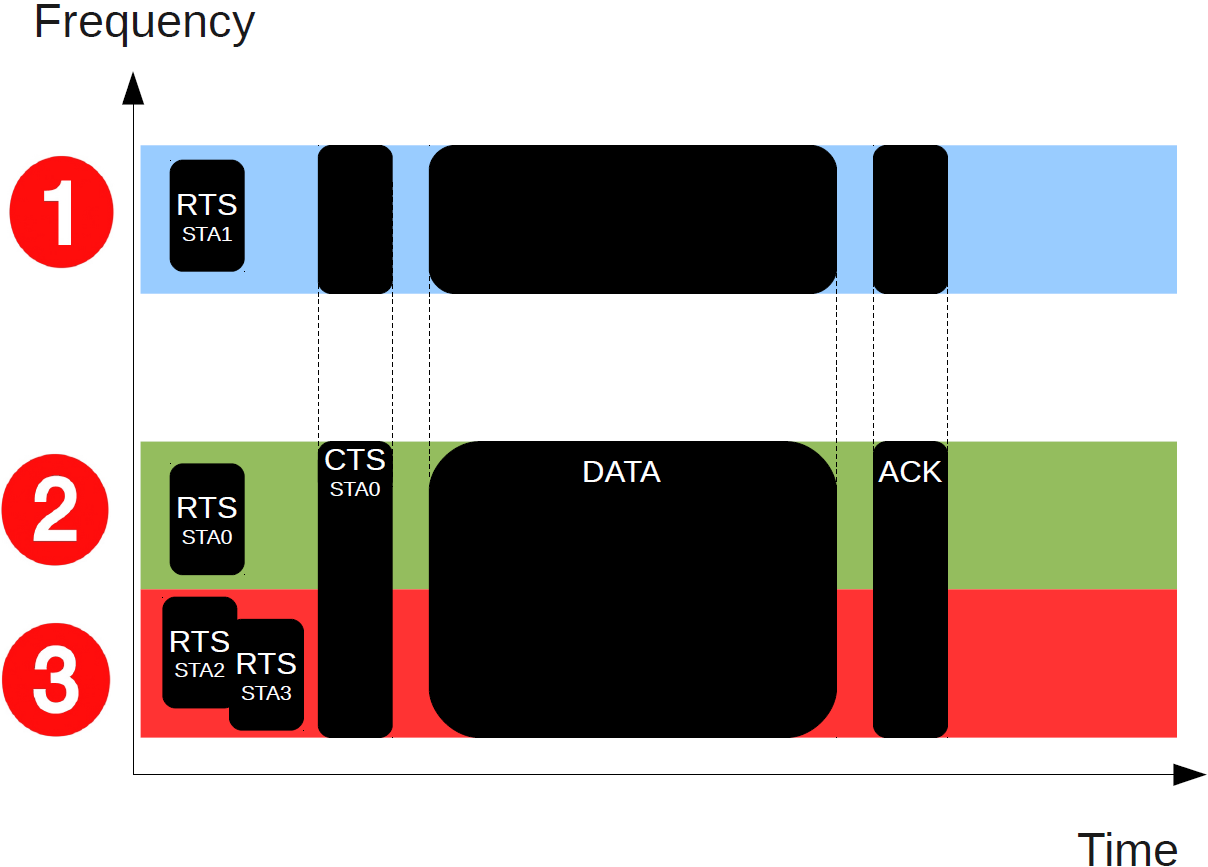}\\
\caption{Multiband CSMA/CA.}
\label{multiband}
\end{center}
\end{figure}

Figure \ref{multiband} provides an exemple with four stations: STA0, STA1, STA2 and STA3, and a single AP. Each STA tries to send an RTS on a band randomly chosen. STA0 and STA1 respectively choose band $2$ and band  $1$ while STA2 and STA3 choose band $3$. At the receiver side a collision occurs on band $3$ but the AP detects both RTS from STA0 and STA1. The AP chooses randomly STA0 and sends CTS over all bands indicating that STA0 has won the channel access. All STAs receive and decode the CTS and only STA0 tries to send its packets during a defined amount of time (several time slots). The communication is said successful when STA0 receives the ACK from the AP.

\section{{Multiband CSMA/CA - RTS/CTS Case Study}}
\label{case_study}
We now explore the benefits and potential issues of this proposed MAC in regards of the classical problems arising with CSMA/CA, such as hidden \cite{hidennodes} and exposed nodes \cite{exposed1} \cite{exposed2}).

\subsection{Hidden node}
The hidden node problem refers to a configuration of three nodes X, R and Y. X can hear R but not Y and Y can hear R but not X. A "hidden node" scenario results when Y attempts to transmit while X is transmitting to R, since Y has sensed the channel idle. The node configuration is depicted in \fig \ref{fig:exposed_node}. This classical problem is resolved by the handshaking mechanism (RTS/CTS). The use of a virtual carrier sense (also known as Network Allocation Vector (NAV) scheme) provides a way to deal with hidden node problem. When a RTS or CTS is received by non transmitting nodes, they defer their backoff during a time specified into the RTS/CTS messages. In the case of the proposed protocol no additional mechanisms are required at the MAC layer. At the physical layer the receiver must be able to analyze each band independently for RTS messages but also to be able to decode the whole band. This is not an issue with OFDM systems.


Last but not least note that if the classical RTS/CTS mechanism avoids collisions in the hidden node scenario, it cannot deal with collisions between RTS messages themselves. The channel is kept clear only when the CTS has been sent. 

\subsection{Exposed node}
\begin{figure}[tb]
\begin{center}
\vspace{-1cm}
\includegraphics[width=1\columnwidth]{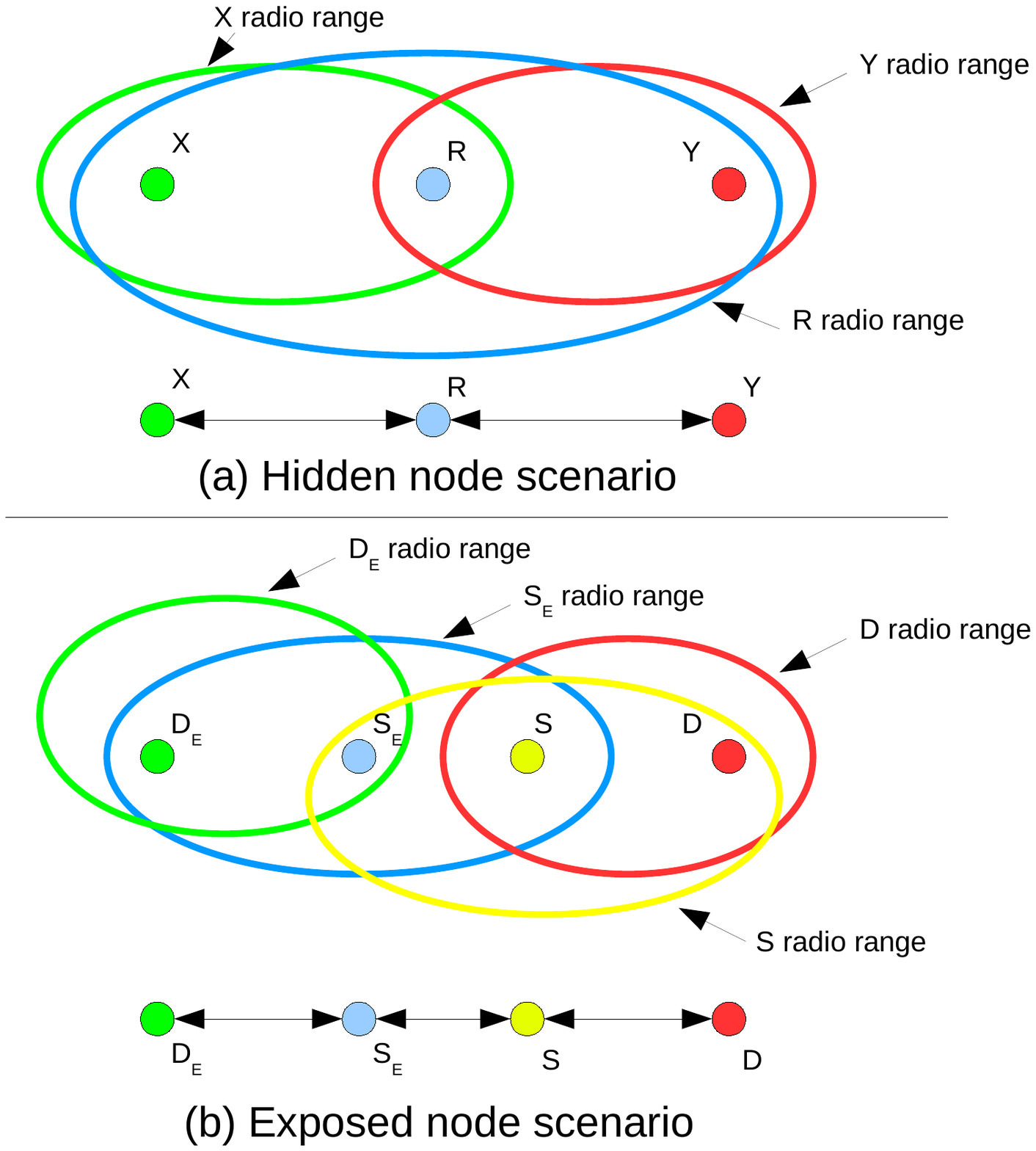}\\
\vspace{-3.5cm}
\caption{Illustration of the hidden and exposed node problem} 
\label{fig:exposed_node}
\end{center}
\end{figure}

RTS/CTS handshake mechanism was introduced to deal with the hidden node issue. However this mechanism introduces a new problem, known as exposed node. The issue of exposed node is depicted in Figure \ref{fig:exposed_node}. Exposed node $S_E$ can hear the RTS and DATA packets sent out from node $S$ to $D$. Consequently, through the virtual carrier sensing, $S_E$ can not initiate transmission despite being out of range of the receiver $D$. Consequently, the transmission between $S_{E}$ and $D_{E}$ is differed introducing a lost in capacity. The same problem exists with the proposed protocol but dealing with this issue is kept out of the scope of this paper. It is worth mentioning that some mechanisms have been proposed in the literature to face the exposed node problem and they could be transposed to the multiband RTS/CTS CSMA/CA protocol (see \cite{exposed1} for instance).

\subsection{New pathologic case}
The frequency multiplexing of RTS introduces a new issue that can be easily solved by a basic rule. Let us consider the following scenario including four nodes, two sources and two destinations. Source $A$ sends a RTS to node $B$ using band $i$ and at the same time, source $C$ sends a RTS to node $D$ using band $i+1$. Node $B$ can hear both $A$ and $C$, while node $D$ can hear $A$ or $C$ only\footnote{when all nodes can hear each other the same problem occurs}. In this case no RTS collision occurs since RTS messages are sent on different bands. Without particular rule the two destinations will respond CTS. In some cases, this scenario can introduce a CTS collision (since CTS messages are broadcasted over all the bands). To prevent the CTS collision and its consequences (watchdog timer is required if no data packet arrives ...), we propose to use the destination identity field already present in the RTS message in order to detect what we call virtual RTS collision. When two or more RTS can be decoded, the destination analyzes the identity of the destination node. If at least two different identities are detected then a collision is declared and no CTS is broadcasted over the cell. This case does not exist in the context of per-AP single band CSMA/CA-RTS/CTS with frequency reuse since each AP send its CTS over its own band. 



\section{Protocol Application}
\label{application}
In this section we discuss a real application to exhibit the motivation for this work. 

We consider an uplink scenario with a random distribution of users  sharing the same bandwidth in a cell using CSMA/CA with a RTS/CTS mechanism. As we know, the signal to noise ratio (SNR) depends on the user position relatively to the access point. 
Each user experience a different SNR, and accordingly a specific capacity. 

\begin{figure}[tb]
\begin{center}
\includegraphics[width=1\columnwidth,height=0.65\columnwidth]{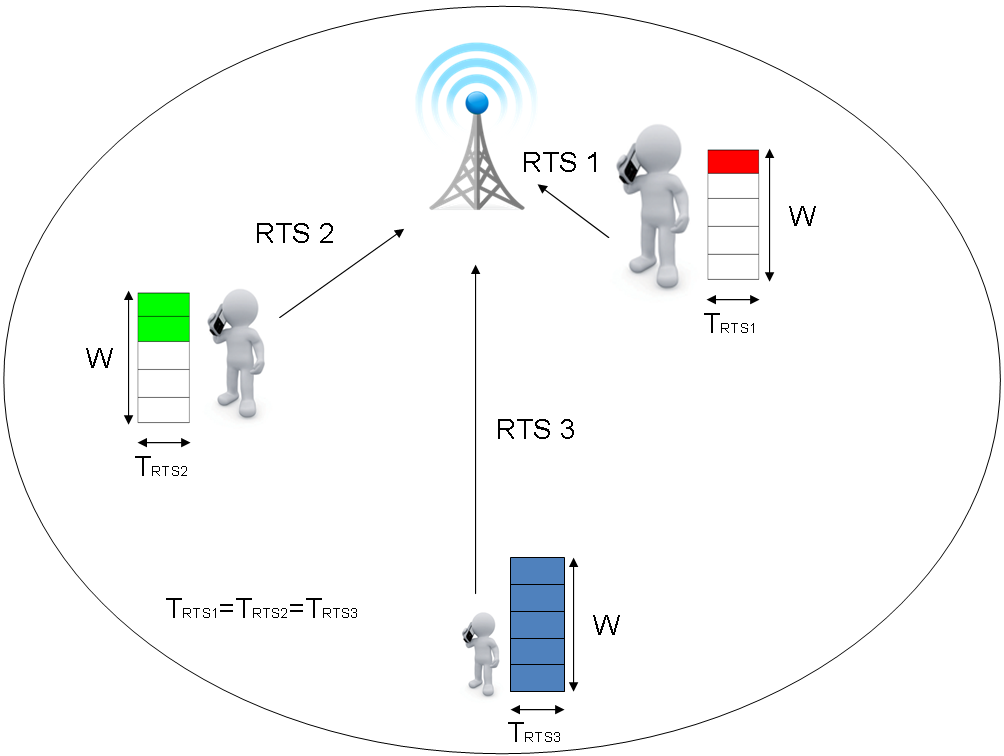}\\
\caption{Multi users with uplink communications.}
\label{scenario}
\end{center}
\end{figure}

Suppose as described in Figure \ref{scenario} that there are three users ready to transmit data (backoff equals zero)  to the AP. User1 is close to the AP, user3 is far from the AP and user2 is in the middle. In order to keep the system working properly, the time of RTS should be the same regardless of the users channel capacity. Thus, all users are penalized by the farthest one since the duration of the RTS should be kept equal to ensure the proper behavior of the protocol. The fact that users with a high SNR does not exploit their whole capacity for RTS represents a spectral efficiency loss. 

But with the multi-bands protocol,  the duration of the RTS can be kept small by allowing distant users to transmit their RTS over several bands.
As an example, Figure \ref{scenario} shows that user3 (highest SNR) uses one band over five and user2 uses two bands over five.\\
For example, if user1 send RTS on the fifth band and user2 send its RTS on first two bands (considering user3 does not transmit), the AP will be able to decode the two messages and choose the qualified user to establish communication. In this case we achieve a successful transmission. One successful RTS transmission (band contains only $1$ RTS) leads to successful communication.

In this context, if user $3$ with the lowest SNR also sends its RTS  to the AP, it can use the whole band for its RTS and the protocol becomes equivalent to the classical one. But high SNR users may take advantage of the multiband protocol. As described in the system model, CTS messages are sent from the AP over all the bands anyway in order to be detectable and decodable by all users regardless their SNR.

\section{Simulation Results}
\label{numerical_results}
In this section, due to the lack of place, we restrict the study to the case where all users have a SNR good enough to be able to use the multiband protocol. 

We focus our study on the impact of the number of RTS bands on the system performance. A home-made event-driven simulator was used to model the protocol behavior. The protocol and channel parameters are reported in Table \ref{parameters} and correspond to those of $802.11n$ standard. The minimal contention window ($W_{min}$) has been chosen constant and equal to $16$. It is worth mentionning that as the study focuses on the MAC mechanisms, an ideal physical layer (no path loss, no fading, no shadowing, ...) is considered.
\begin{table}[htb]
\centering
\begin{tabular}{|l|c|c|}
\hline
Packet payload & 8184 bits\\
MAC header & 272 bits \\
PHY header & 128 bits \\
ACK length & 112 bits + PHY header \\
RTS length & 160 bits + PHY header \\
CTS length & 112 bits + PHY header \\
Channel Bit Rate & 72.2 Mbit/s \\
Propagation Delay & 1 $\mu$s \\
SIFS & 10 $\mu$s \\
Slot Time & 9 $\mu$s\\
DIFS & 28 $\mu$s\\
\hline
\end{tabular}
\caption{PHY layer parameters for 802.11n}
\label{parameters}
\end{table}
\subsection{Collision Probability}
\begin{figure}[tb]
\begin{center}
\includegraphics[width=1\columnwidth,height=0.65\columnwidth]{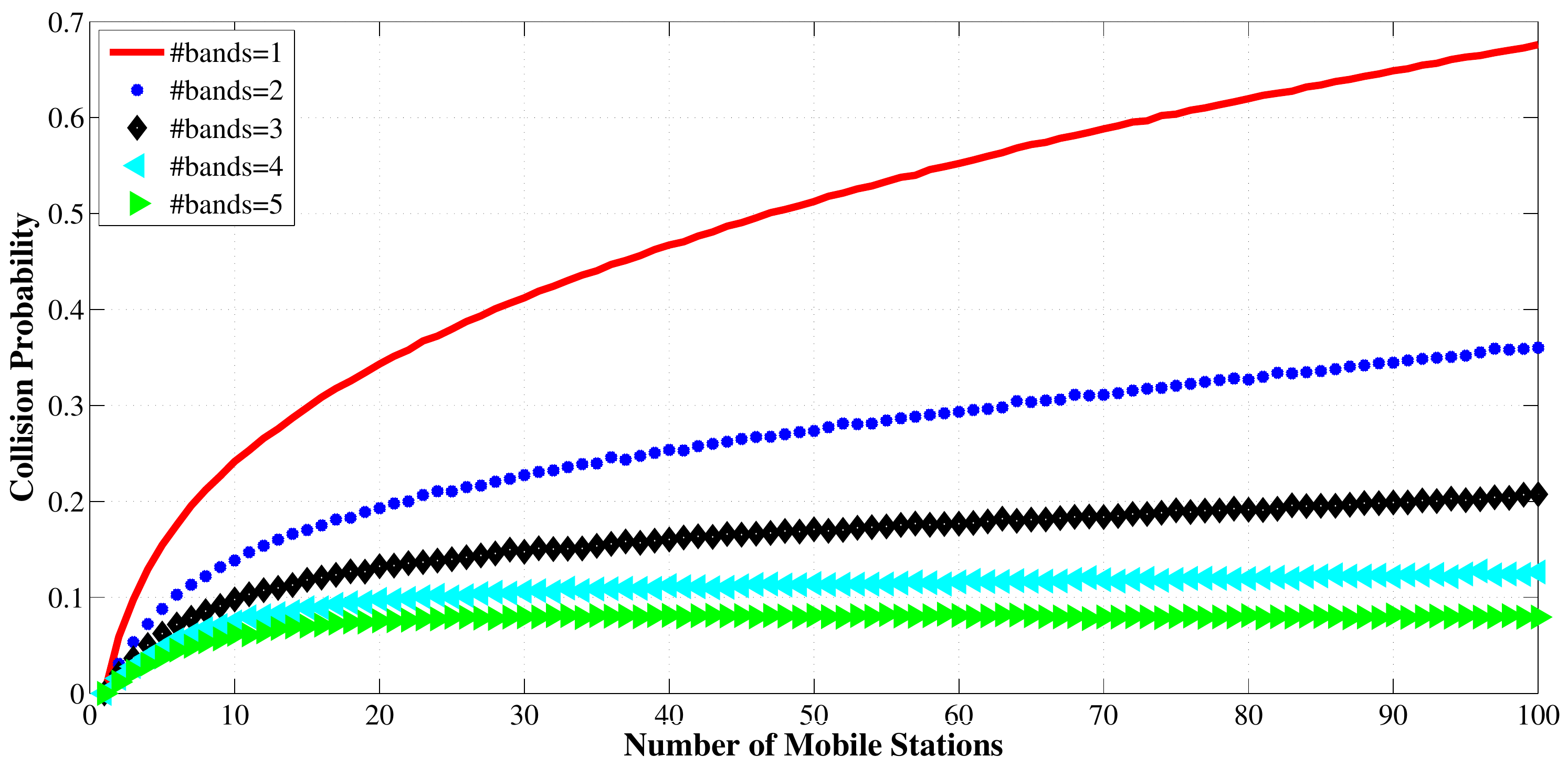}\\
\caption{Collision probability for Multi RTS bands.}
\label{collision}
\end{center}
\end{figure}
\begin{figure}[tb]
\begin{center}
\includegraphics[width=1\columnwidth,height=0.65\columnwidth]{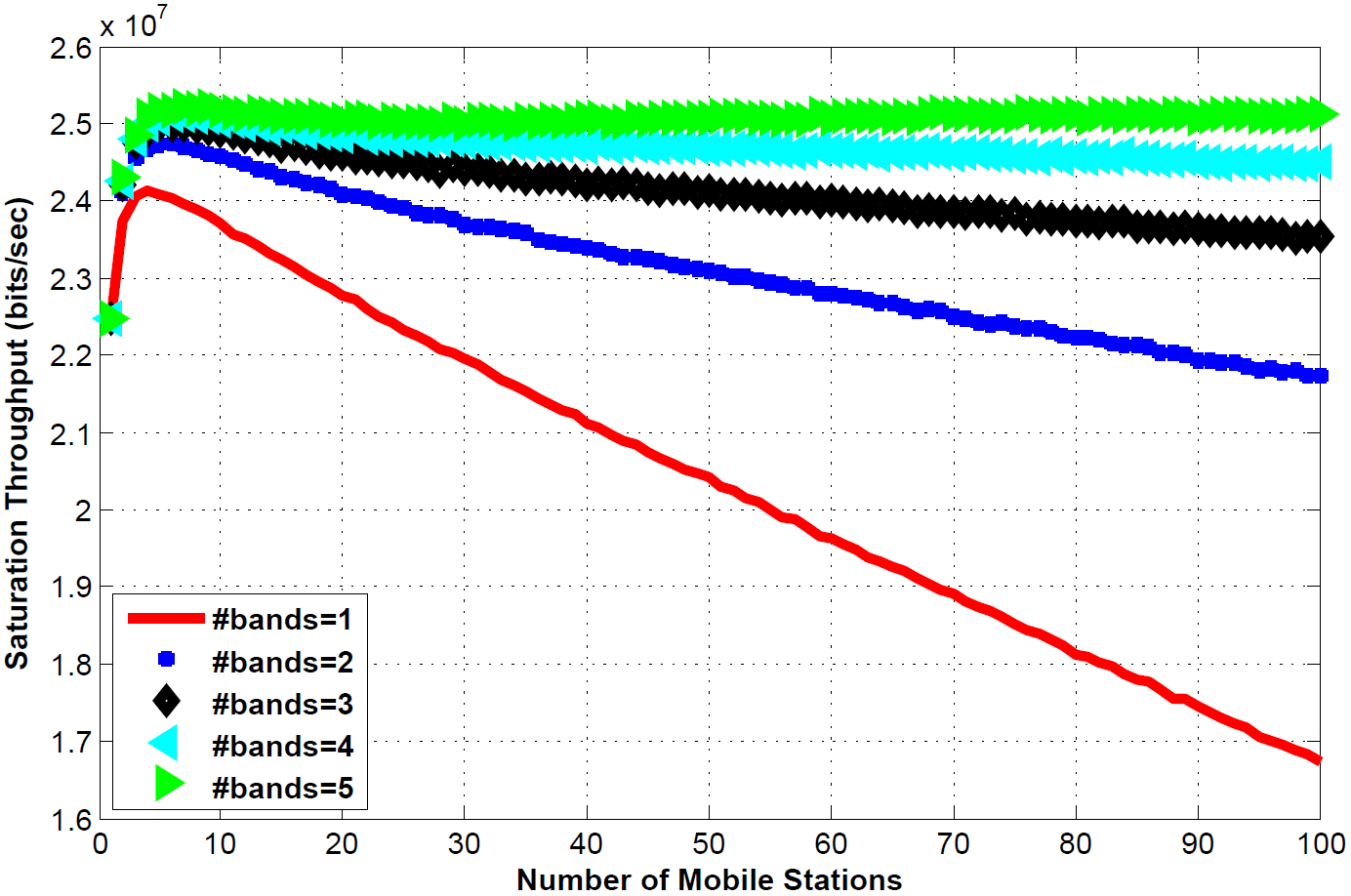}\\
\caption{Saturation throughput (bits/sec) vs. number of mobile stations considering multi RTS bands.}
\label{multiband_chs}
\end{center}
\end{figure}

As the system performance is related to RTS collision probability, it is interesting to study the impact of the proposed band division. We consider different number of sources trying to access a common destination.
Figure \ref{collision} depicts the simulation results for the collision probability between RTS messages as a function of the number of mobile stations present in the network for various RTS bands values.

These results demonstrate that the collision probability increases with the number of users but is inversely proportional to the number of RTS bands. For a single band CSMA/CA with 50 users, the probability of collision is around 50\%. For a two bands protocol the probability of collision is reduced to 25\%. When 5 bands are considered the probability of collision is less than 10\%. As we discussed before, the proposed protocol reduces drastically the RTS collision probability.

As collisions happen only during RTS transmissions (considering perfect channel conditions), the proposed MAC improves the global system performance in terms of throughput and latency.

\subsection{Saturation Throughput}
In this sub-Section we study the throughput in saturation mode, so we suppose that each station has always in its buffer at least one packet ready for transmission. Figure \ref{multiband_chs} depicts the saturation throughput as a function of the number of mobile stations present in the network for various RTS bands values.

It shows that increasing the number of RTS bands in the system improves as well the saturation throughput. Global system performance is improved by having the possibility to detect simultaneous RTS even if the system can deal with only one RTS. This is due to the reduction of the RTS collision probability.

The improvement is significant for low and high number of users. 
Table \ref{throughput} illustrates the gain introduced in the multiband context. It is demonstrated that the gain becomes more important in loaded networks. This protocol brings more than 50\% of gain (comparing multiband to single band in terms of saturation throughput) when the number of RTS bands exceeds four in charged mode (loaded network).

\begin{table}[htb]
\centering
\begin{tabular}{|l|c|c|c|}
\hline
\#Stations & \#RTS Bands & \specialcell{ Saturation Throughput \\ (Mbits/sec)} & Gain (\%)\\
\hline
10 & 2 & 24.56 & 3.57\\
10 & 3 & 24.90 & 5.00\\
10 & 4 & 25.05 & 5.64\\
10 & 5 & 25.17 & 6.12\\
50 & 2 & 23.08 & 13.09\\
50 & 3 & 24.13 & 18.22\\
50 & 4 & 24.66 & 20.84\\
50 & 5 & 25.06 & 22.77\\
100 & 2 & 21.73 & 29.84\\
100 & 3 & 23.53 & 40.56\\
100 & 4 & 24.51 & 46.42\\
100 & 5 & 25.11 & 50.04\\
\hline
\end{tabular}
\caption{Saturation throughput gain with the proposed MAC for different mobile stations and RTS bands number.}
\label{throughput}
\end{table}
\subsection{Statistical Delay Study}
\begin{figure}[tb]
\begin{center}
\includegraphics[width=1\columnwidth,height=0.65\columnwidth]{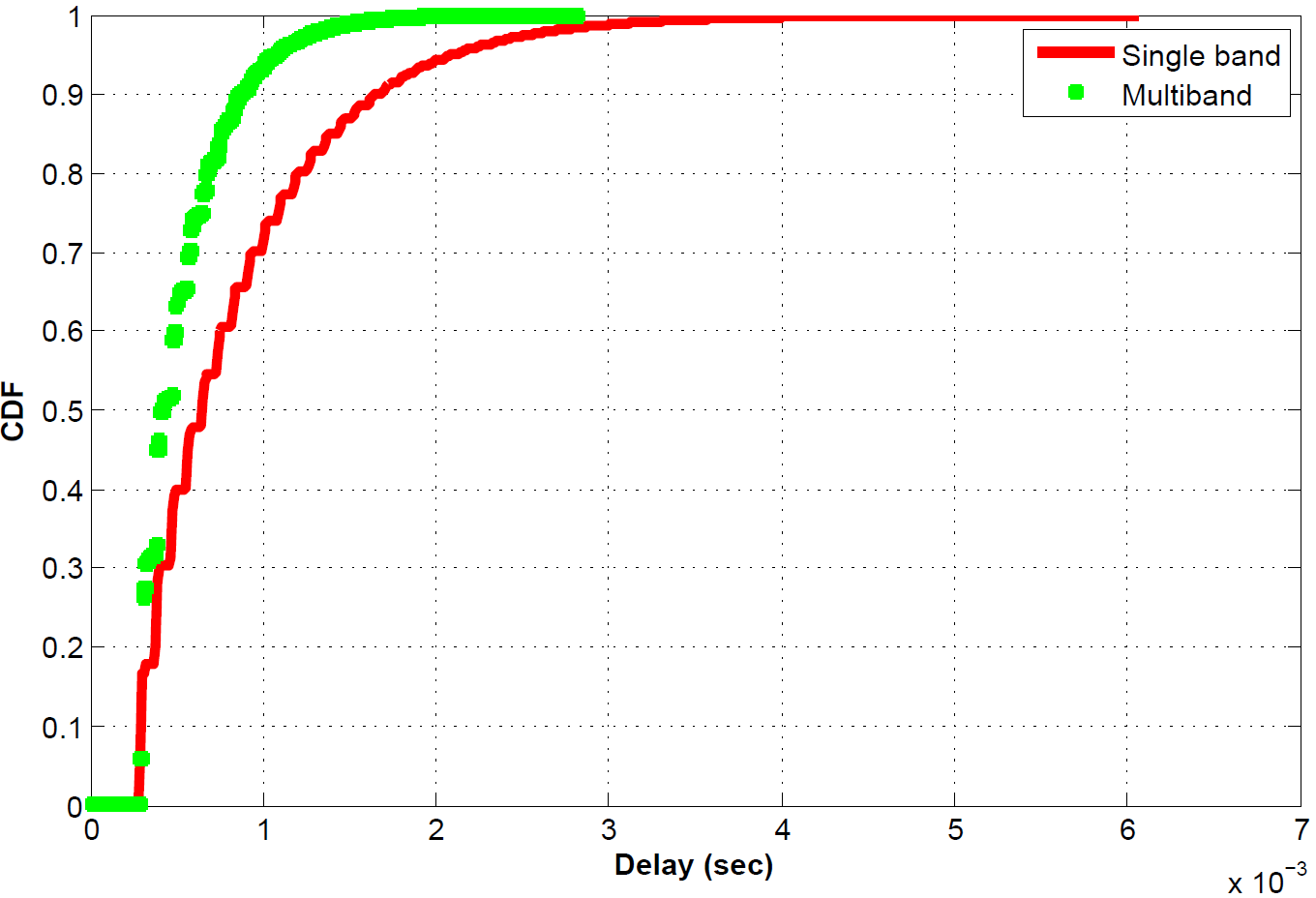}\\
\caption{CDF of access delay with 100 stations for single and multi RTS band. Delay is expressed in second.}
\label{delay}
\end{center}
\end{figure}

\begin{table}[htb]
\centering
\begin{tabular}{|l|c|c|c|}
\hline
CDF & \#RTS Bands & Proposed Protocol Delay $(ms)$ & Gain (\%)\\
\hline
99\% & 2 & 1.83 & 69.73\\
99\% & 3 & 1.61 & 94.46\\
99\% & 4 & 1.53 & 104.65\\
99\% & 5 & 1.48 & 109.61\\
98\% & 2 & 1.62 & 65.29\\
98\% & 3 & 1.39 & 93.72\\
98\% & 4 & 1.33 & 102.19\\
98\% & 5 & 1.30 & 105.15\\
95\% & 2 & 1.28 & 62.35\\
95\% & 3 & 1.13 & 85.44\\
95\% & 4 & 1.09 & 92.00\\
95\% & 5 & 1.05 & 97.61\\
90\% & 2 & 1.02 & 61.98\\
90\% & 3 & 0.92 & 78.45\\
90\% & 4 & 0.87 & 88.34\\
90\% & 5 & 0.86 & 89.21\\
\hline
\end{tabular}
\caption{Delay gain with the proposed MAC for different rts bands number with 100 users.}
\label{delay_gain}
\end{table}

To complete the study we go forward to simulate the delay introduced by the proposed MAC protocol. The delay is defined as the duration needed to transmit a packet. In order to compare the delay between the two strategies (single and multiband), we extract from simulation the cumulative density function (CDF) of the delay for one network scenario and for many number of users. Figure \ref{delay} illustrates that the CDF of the access delay is better for the multiband scheme. For instance, 99\% of packets are transmitted with at most $3.13 ms$ by the single band protocol while they are sent with at most $1.53 ms$ by our proposed MAC protocol with $4$ RTS bands.

Different gain values introduced by the proposed MAC are reported in table \ref{delay_gain}. The gain is computed by comparing both single and multiband CDF. As seen in Table \ref{delay_gain}, using multiband protocol, the delay is reduced by half in loaded networks (considering more than $3$ RTS bands). This improvement is explained by the fact that the proposed protocol reduces the collision probability between RTS, hence packets wait less before to be transmitted.

\begin{figure}[tb]
\begin{center}
\includegraphics[width=1\columnwidth]{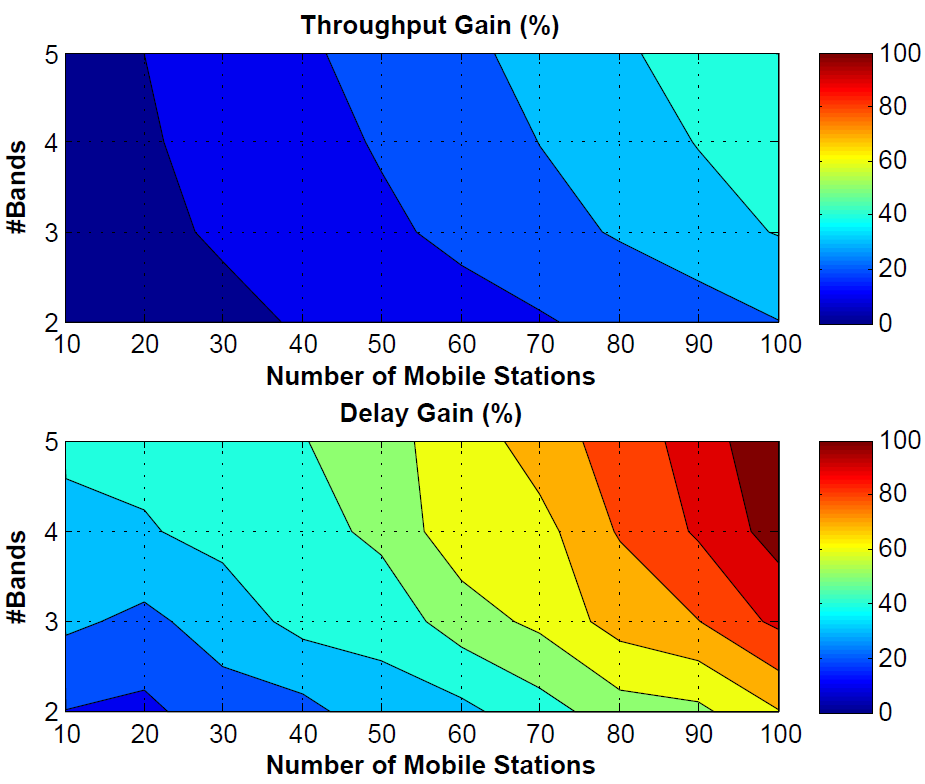}\\
\caption{Saturation throughput and Delay vs. number of mobile stations considering various number of RTS bands.}
\label{synthese}
\end{center}
\end{figure}

Figure \ref{synthese} depicts the saturation throughput and delay gains (\%) vs. the number of RTS bands and mobile stations. It should be noticed that the gain in terms of saturation throughput and delay are always positives and becomes much important in the case of loaded networks. Increasing the number of RTS bands improves the system performance (as the RTS collision probability is reduced). 
Moreover, this protocol improves the system latency since the collision probability and at the same time the number of contented users reduce.

\section{Conclusion}
\label{conclusion}
In this paper, we proposed an innovative scheme exploiting a random frequency division multiplexing of RTS messages in a CSMA/CA RTS/CTS access method.

This technique is characterized by considering a spectrum which is divided into several bands of known size. We demonstrated that the proposed MAC is very interesting especially in crowded networks. By considering a frequency division multiplexing of RTS messages, the probability of RTS collisions is decreased significantly. We achieved a gain of about $50\%$ in terms of saturation throughput and $109\%$ in terms of delay. Due to these good properties in crowded scenario, the proposed protocol is also a good candidate for wireless Machine to Machine (M2M) applications in which latency is critical.


\bibliographystyle{IEEEtran}
\bibliography{bibliography}

\end{document}